\documentclass{JHEP3}
\usepackage{amsmath}
\usepackage{times}
\usepackage[mathscr]{eucal}

\newcommand{\bi}{{\bar\imath}}
\newcommand{\bj}{{\bar\jmath}}

\title{Supersymmetric Gauge Theories on the Five-Sphere}

\author{
Kazuo Hosomichi$^{\dagger a}$,
Rak-Kyeong Seong$^{* b}$ 
and
Seiji Terashima$^{\dagger c}$\\
${}^\dagger$Yukawa Institute for Theoretical Physics,\\
~  Kyoto University, Japan\\
${}^* $Theoretical Physics Group, The Blackett Laboratory,
Imperial College London,\\
Prince Consort Road, London SW7 2AZ, UK\\
${}^a$ E-mail:
\email{hosomiti@yukawa.kyoto-u.ac.jp}\\
${}^b$ E-mail:
\email{rak-kyeong.seong@imperial.ac.uk}\\
${}^c$ E-mail:
\email{terasima@yukawa.kyoto-u.ac.jp}\\
}

\preprint{YITP-12-10, Imperial/TP/12/RS/02}

\abstract{
We construct Euclidean 5d supersymmetric gauge
theories on the five-sphere with vector and hypermultiplets.
The SUSY transformation and the action are explicitly determined 
from the standard Noether procedure as well as from off-shell supergravity.
Using localization techniques, the path-integral is shown to be
restricted to the integration over
a generalization of
instantons on $\mathbb{CP}^2$ and the Coulomb moduli.
}

\keywords{Supersymmetric gauge theory}

\begin{document}

\section{Introduction}\label{sec:intro}

Recently, supersymmetric (SUSY) gauge theories on curved spaces 
have been studied intensively. 
One of the reasons why such studies are important is 
that the partition function and some correlators 
are exactly computable by localization techniques.
In general, however, it is not obvious whether one can construct
SUSY gauge theories on a given curved space. Even if it is possible, one has to construct the theories one by one.

SUSY gauge theories on some simple curved spaces
have caught attention for many years.\footnote{
There have been constructions of SUSY gauge theories on curved
spaces based on topological twisting \cite{Witten, Vafa-Witten}.
}
One of the simplest classes of SUSY gauge theories on curves spaces are on $d$-dimensional spheres $S^d$.
SUSY gauge theories on $S^4$
were considered in \cite{Pestun},
where it was shown that the partition function
as well as the expectation values of certain Wilson loops
can be computed as certain matrix integrals
which were first conjectured by \cite{Erickson-SZ, Drukker-G}.
The exact results also led to the discovery of a surprising relation
between 4d ${\cal N}=2$ SUSY gauge theories and 2d conformal field
theories \cite{AGT, Wyllard}.
For three dimensions, an exact computation was initiated by
\cite{Kapustin-WY1} and generalized by \cite{Jafferis, Hama-HL}.

Following these successes in finding SUSY theories on $S^4$ and $S^3$, 
it is natural to extend the search and to consider SUSY gauge theories on $S^5$.
Although 5d gauge theories are not perturbatively renormalizable,
one can consider any UV completion of the theory on $S^5$.
Then, if localization gives an exact result
which is independent of how one completes the theory in the UV,
the result may be well-defined.

Adding to this,
there is another strong motivation to consider these theories.
In \cite{Douglas, Lambert-PS}, it was proposed that
the maximal SUSY 5d gauge theory describes
the 6d $\mathcal{N}=(2,0)$ SUSY conformal field theory compactified
on a circle without introducing Kaluza-Klein degrees of freedom.
This 6d $\mathcal{N}=(2,0)$ CFT is both very interesting and mysterious.
There is no intrinsic definition of the theory,
and we only know of it through
(non-perturbatively defined) string theory or M-theory.
It is therefore important to check this highly non-trivial proposal.
For instance, the gravity dual analysis implies that
there should be ${\cal O}(N^3)$ degrees of freedom at large $N$,
instead of $N^2$ as it is naively expected from a 5d gauge theory.
So far, there are some indications 
which support the proposal. These indications are from, for example, 
the analysis of the M2-M5 bound state \cite{Terashima-Yagi, Lambert-N}
and the instanton index counting in flat 5d \cite{Kim-K}. But despite these indications, there has not yet been enough evidence for the proposal.
The exact partition function for 5d SUSY gauge theories on $S^5$
will serve as a more direct measure of the degrees of freedom,
like the 3d counterpart which was successfully applied
to the case of multiple M2-branes \cite{Drukker-MP1,Drukker-MP2, Marino-P}. 

In this paper, 
we construct the (Euclidean) 
five-dimensional $\mathcal{N}=1$ SUSY gauge theory on $S^5$
with vector and hypermultiplets.
This is not a conformal field theory, thus there are 8 SUSY
generators and the $SU(2)$ R-symmetry is broken
to $U(1)$ by the curvature of $S^5$.
These include the analogue of the $\mathcal{N}=2$ SUSY 5d gauge theory on
${\mathbb R}^5$
for the one on $S^5$, i.e. a vector multiplet and 
an adjoint hypermultiplet.
The SUSY transformation and the action are determined 
from the standard Noether procedure.
We also show that, as suggested by \cite{Festuccia-S},
the rigid SUSY transformation and the action can be obtained
from the off-shell $(4+1)$d supergravity theory \cite{Kugo-Ohashi, Kugo-Ohashi2}
by choosing the VEV of the fields in the supergravity multiplet.
Indeed, by an appropriate choice of the auxiliary field
we show that this also gives the 
same 5d SUSY gauge theory on $S^5$ for the vector multiplet.

In order to apply the localization technique,
one needs to choose a SUSY generator, which is 
equivalent to choosing a $SU(2)_R$ doublet spinor.
We show that there is essentially only one choice.
The bi-linear of the spinors 
is a vector field without fixed points, which
leads to a $S^1$ fibration over $\mathbb{CP}^2$.
Then, for the vector multiplet
the standard term for localization \cite{Pestun}
restricts the path-integral to an integration over
a generalization of instantons 
on $\mathbb{CP}^2$ and the covariantly constant Coulomb moduli.
Unfortunately, 
we have not succeeded in evaluating the 
localized path-integral for now, but hope to return to this problem in the near future.

The organization of this paper is as follows:
section 2 gives a short review of the 5d spinor calculus
and then constructs the 5d SUSY gauge theory on $S^5$
with vector multiplets and hypermultiplets.
Explicit Killing spinors are given at the end of the section.
In section 3, localization is applied to 
the 5d gauge theories on $S^5$. 
We conclude with a short discussion in section 4.
\\

\section{5D SUSY gauge theory on $S^5$}\label{sec:5D SUSY gauge theory on $S^5$}

The 5d SUSY gauge theory on $S^5$ is constructed in this section.

\subsection{5D Spinor Calculus}\label{subsec:spinor}

First, we summarize the properties of the Euclidean 5d spinors
in ${\mathbb R}^5$.
The 5d Gamma matrices are a set of $4\times 4$ hermitian matrices satisfying
$\left\{\Gamma^m,\Gamma^n\right\}=2\delta^{mn}$. The standard
notation for their antisymmetrized products is
\[
\Gamma^{n_1n_2\cdots n_p}~\equiv~\frac1{p!}
\big(\Gamma^{n_1}\Gamma^{n_2}\cdots\Gamma^{n_p}\pm\cdots\big)~~.
\]
We also need a matrix $C$ which relates $\Gamma^m$ to its transpose as
$C\Gamma^mC^{-1}=\pm(\Gamma^m)^T$, called the charge conjugation matrix.
Assuming $C$ to be either symmetric or antisymmetric, one easily finds that
$C\Gamma^{n_1\cdots n_p}$ all have definite parity under transposition.
Using
\[
\Gamma^{n_1n_2n_3n_4n_5}~=~ \varepsilon^{n_1n_2n_3n_4n_5}~~,
\]
one can argue that $C$ and $C\Gamma^m$ are antisymmertic whereas
$C\Gamma^{mn}$ are symmetric. They span the 16-dimensional linear
space of $4\times 4$ matrices. Accordingly, one also has
\[
C\Gamma^mC^{-1}=(\Gamma^m)^T=(\Gamma^m)^\ast~~.
\]
The above further implies that $(C^\ast C)\Gamma^m(C^\ast C)^{-1}=\Gamma^m$
and that $C^\ast C=-C^\dag C$ is an Hermitian matrix which commutes
with any $4\times 4$ matrix; i.e. it is proportional to the identity.
We normalize with
\[
C^\ast C=-1~~.
\]
In our convention the matrices have the following index structure
\begin{equation}
C_{\alpha\beta}~~,\quad
(\Gamma^m)^\alpha_{~~\beta}~~,\quad
(C\Gamma^{n_1\cdots n_p})_{\alpha\beta}~~,\quad
-(C^\ast)^{\alpha\beta}=(C^{-1})^{\alpha\beta}\equiv C^{\alpha\beta}~~.
\end{equation}

Spinors $\psi^\alpha$ belong to a 4-dimensional representation of the
rotation group $Sp(2)\simeq SO(5)$. This representation is pseudoreal, and $(\psi^\ast)_\alpha$ and $(C\psi)_\alpha\equiv
C_{\alpha\beta}\psi^\beta$ transform the same way under rotations.
Because it is the pseudoreal representation,
the Majorana (or real) condition $\psi^\ast\equiv C\psi$ does not satisfy
the consistency $\ast\ast=\text{id}$ due to $C^\ast C=-1$. 

The 5d SUSY algebra has $SU(2)$ R-symmetry as an automorphism.
For spinors which are doublets under $SU(2)_\text{R}$,
one can instead impose the $SU(2)$ Majorana condition\footnote{
Here we are considering Euclidean instead of Lorentzian signature, 
but we still call this $SU(2)$ Majorana. Note that 
in the Lorentzian signature there are also $SU(2)$ Majorana spinors.},
\begin{equation}
( \psi^\alpha_I)^\ast ~=~ \epsilon^{IJ}C_{\alpha\beta}\psi^\beta_J~~.
\end{equation}
since they are in real(-positive) representations of $SU(2)
\times Spin(5)$.
Here, $\epsilon^{IJ}$ is the antisymmetric $SU(2)$ invariant tensor
defined by $\epsilon^{12}=-\epsilon^{21}=1$. We also introduce
$\epsilon_{IJ}$ with $\epsilon_{12}=-\epsilon_{21}=-1$.

For bilinear spinors, we use the following notation
\begin{equation}
\xi\eta \equiv C_{\alpha\beta}\xi^\alpha\eta^\beta~~,\quad
\xi\Gamma^{n_1\cdots n_p}\eta\equiv
(C\Gamma^{n_1\cdots n_p})_{\alpha\beta}\xi^\alpha\eta^\beta~~.
\end{equation}
\\

\subsection{Vector Multiplets}\label{subsec:vector}

In this subsection, we concentrate on vector multiplets
for an arbitrary gauge group.
We first review the 5d SUSY gauge theory 
on flat ${\mathbb R}^5$ with off-shell component fields 
and
then study the theory on $S^5$.
\vspace{0.2cm}

\paragraph{Flat $\mathbb R^5$.}
A vector multiplet contains a 5d vector $A_m$, a real scalar $\sigma$,
a triplet of auxiliary scalars $D_{IJ}$ satisfying $
(D_{IJ})^\dag 
=
D^{IJ}
\equiv
\epsilon^{II'}\epsilon^{JJ'}D_{JJ'}$, and an $SU(2)$
Majorana spinor $\lambda_I^\alpha$. For non-abelian gauge symmetry, we
assume these fields are Hermitian matrix-valued. On flat $\mathbb R^5$,
their SUSY variation takes the form
\begin{eqnarray}
\delta_\xi A_m &=& i\epsilon^{IJ}\xi_I\Gamma_m\lambda_J~~, \nonumber \\
\delta_\xi\sigma &=& i\epsilon^{IJ}\xi_I\lambda_J~~,\nonumber \\
\delta_\xi\lambda_I &=&
-\frac12\Gamma^{mn}\xi_IF_{mn}+\Gamma^m\xi_ID_m\sigma
+\xi_JD_{KI}\epsilon^{JK}~~, \nonumber \\
\delta_\xi D_{IJ} &=&
-i(\xi_I\Gamma^mD_m\lambda_J+\xi_J\Gamma^mD_m\lambda_I)
+[\sigma,\xi_I\lambda_J+\xi_J\lambda_I]~~,
\end{eqnarray}
where we used
\begin{eqnarray}
F_{mn} &=& \partial_mA_n-\partial_nA_m-i[A_m,A_n]~~, \nonumber \\
D_m\sigma &=& \partial_m\sigma-i[A_m,\sigma]~~.
\end{eqnarray}
These transformation laws are consistent with Hermite conjugation
as one can check by using $(\Gamma^m)^\ast=C\Gamma^mC^{-1}$ along with
\begin{equation}
(\xi_I\lambda_J)^\dag =
-\epsilon^{II'}\epsilon^{JJ'}\xi_{I'}\lambda_{J'}~~,\quad
(\xi_I\Gamma_m\lambda_J)^\dag =
-\epsilon^{II'}\epsilon^{JJ'}\xi_{I'}\Gamma_m\lambda_{J'}~~,
\end{equation}
\begin{equation}
(\epsilon^{IJ}\xi_I\lambda_J)^\dag =
-\epsilon^{IJ}\xi_I\lambda_J~~,\quad
(\epsilon^{IJ}\xi_I\Gamma_m\lambda_J)^\dag =
-\epsilon^{IJ}\xi_I\Gamma_m\lambda_J~~,
\end{equation}
where we assumed that $\xi_I$ is an $SU(2)$ Majorana fermion.
The coefficients of various terms are determined by requiring that the
commutator of two SUSY yields the Lie derivative ${\cal L}(-i v)$ and gauge
transformation ${\cal G}$,
\begin{equation}
[\delta_\xi,\delta_\eta]~=~ {\cal L}(-i v)+{\cal G}(\gamma+i v^mA_m)~~,
\end{equation}
with
\begin{equation}
v^m = 2 \epsilon^{IJ}\xi_I\Gamma^m\eta_J~~, \quad
\gamma = -2i\epsilon^{IJ}\xi_I\eta_J\sigma~~.
\end{equation}
More explicitly,
\begin{eqnarray}
~[\delta_\xi,\delta_\eta]A_m &=& -i v^nF_{nm} + D_m\gamma~~, \nonumber \\
~[\delta_\xi,\delta_\eta]\sigma &=& -i v^nD_n\sigma~~,
\nonumber \\
~[\delta_\xi,\delta_\eta]\lambda_I &=&
-i v^nD_n\lambda_I+i[\gamma,\lambda_I]~~, \nonumber \\
~[\delta_\xi,\delta_\eta]D_{IJ} &=&
-i v^nD_nD_{IJ}+i[\gamma,D_{IJ}]~~.
\end{eqnarray}
In order to derive the above, one needs make use of the Fierz identity which holds for any
three spinors $(\xi,\eta,\psi)$\footnote{
This is valid if $(\xi,\eta,\psi)$ are fermions. For bosonic spinors,
we have
\begin{equation}
\xi^\alpha (\eta\psi) ~=~
\frac14\psi^\alpha (\eta\xi)
+\frac14(\Gamma^m\psi)^\alpha (\eta\Gamma_m\xi)
-\frac18(\Gamma^{mn}\psi)^\alpha (\eta\Gamma_{mn}\xi)~.
\end{equation}},
\begin{equation}
\xi^\alpha (\eta\psi) ~=~
-\frac14\psi^\alpha (\eta\xi)
-\frac14(\Gamma^m\psi)^\alpha (\eta\Gamma_m\xi)
+\frac18(\Gamma^{mn}\psi)^\alpha (\eta\Gamma_{mn}\xi)~~.
\end{equation}
Accordingly, the Yang-Mills term $\frac12\text{tr}(F_{mn}F^{mn})$ has the following
SUSY completion
\begin{equation}
{\cal L}_\text{SYM} =
\text{tr}\left[
\frac12F_{mn}F^{mn}-D_m\sigma D^m\sigma-\frac12D_{IJ}D^{IJ}
+i\epsilon^{IJ}\lambda_I\Gamma^mD_m\lambda_J
-\epsilon^{IJ}\lambda_I[\sigma,\lambda_J]
\right]~~.
\end{equation}
\vspace{0.2cm}

\paragraph{Five-Sphere.}

For SUSY theories on $S^5$, the supersymmetry transformation parameter
$\xi_I$ is expected to be a Killing spinor satisfying
\begin{equation}
D_m\xi_I \equiv
\left(\partial_m+\frac14\omega_m^{ab}\Gamma^{ab}\right)\xi_I =
\Gamma_m\tilde\xi_I
\end{equation}
with a certain $\tilde\xi_I$ from the 4d and 3d computations
\cite{Pestun} \cite{Kapustin-WY1}. 
Here $D_m$ is the local Lorentz covariant
derivative and $\omega^{ab}_m$ is the spin connection. We also need to
distinguish the curved indices ($m,n,\cdots$) and flat indices
($a,b,\cdots$). $\Gamma^a$ is constant but $\Gamma_m=e_m^a\Gamma^a$ is
coordinate-dependent.
We will show that $\tilde\xi$ will be given by $\xi$.
In section \ref{s5killing}, 
we explicitly construct 
the Killing spinors on $S^5$.

The SUSY variation of fields on $S^5$ takes the form
\begin{eqnarray}
\delta_\xi A_m &=& i\epsilon^{IJ}\xi_I\Gamma_m\lambda_J~~, \nonumber \\
\delta_\xi\sigma &=& i\epsilon^{IJ}\xi_I\lambda_J~~,\nonumber \\
\delta_\xi\lambda_I &=&
-\frac12\Gamma^{mn}\xi_IF_{mn}+\Gamma^m\xi_ID_m\sigma
+\xi_JD_{KI}\epsilon^{JK}+2\tilde\xi_I\sigma~~, \nonumber \\
\delta_\xi D_{IJ} &=&
-i(\xi_I\Gamma^mD_m\lambda_J+\xi_J\Gamma^mD_m\lambda_I)
+[\sigma,\xi_I\lambda_J+\xi_J\lambda_I]
+i(\tilde\xi_I\lambda_J+\tilde\xi_J\lambda_I)~~.
\nonumber\\
\label{susytr}
\end{eqnarray}
This form is determined from the requirement that the commutator of two
SUSY should be a sum of translation ($v^m$), gauge transformation
($\gamma+i v^m A_m$), 
dilation ($\rho$), R-rotation ($R_{IJ}$) and Lorentz rotation ($\Theta^{ab}$):
\begin{eqnarray}
~[\delta_\xi,\delta_\eta]A_m &=& -i v^nF_{nm} + D_m\gamma~~, \nonumber \\
~[\delta_\xi,\delta_\eta]\sigma &=& -i v^nD_n\sigma 
+\rho\sigma~~, \nonumber \\
~[\delta_\xi,\delta_\eta]\lambda_I &=&
-i v^nD_n\lambda_I+i[\gamma,\lambda_I]+\frac32\rho\lambda_I
+R_I^{~J}\lambda_J
+\frac14\Theta^{ab}\Gamma^{ab}\lambda~~, \nonumber \\
~[\delta_\xi,\delta_\eta]D_{IJ} &=&
-i v^nD_nD_{IJ}+i[\gamma,D_{IJ}]+2\rho D_{IJ}
+R_I^{~K}D_{KJ}+R_J^{~K}D_{IK}
\nonumber \\ && -2i\sigma(
\eta_I\Gamma^mD_m\tilde\xi_J+\eta_J\Gamma^mD_m\tilde\xi_I
-\xi_I\Gamma^mD_m\tilde\eta_J-\xi_J\Gamma^mD_m\tilde\eta_I )~~.
\label{com}
\end{eqnarray}
Here $R_I^{~J}=\epsilon^{JK}R_{IK}$ and
\begin{eqnarray}
v^m &=& 2\epsilon^{IJ}\xi_I\Gamma^m\eta_J~~, \nonumber \\
\gamma &=& -2i\epsilon^{IJ}\xi_I\eta_J\sigma~~, \nonumber \\
\rho &=& -2i\epsilon^{IJ}(\xi_I\tilde\eta_J-\eta_I\tilde\xi_J)~~,
\nonumber \\
R_{IJ} &=& -3i(\xi_I\tilde\eta_J+\xi_J\tilde\eta_I
             -\eta_I\tilde\xi_J-\eta_J\tilde\xi_I)~~, \nonumber \\
\Theta^{ab} &=&
-2i\epsilon^{IJ}(\tilde\xi_I\Gamma^{ab}\eta_J-\tilde\eta_I\Gamma^{ab}\xi_J)~~.
\label{intsym}
\end{eqnarray}
The unwanted last term in the right hand side of
$[\delta_\xi,\delta_\eta]D_{IJ}$ vanishes if we require
\begin{equation}
\Gamma^mD_m\tilde\xi_I=h\cdot\xi_I\quad\text{or equivalently}\quad
\Gamma^mD_m\Gamma^nD_n\xi_I = 5h\cdot\xi_I
\end{equation}
for a certain scalar function $h$. Note that this also implies
\begin{equation}
\Gamma^{mn}D_mD_n\xi_I
= \frac18\Gamma^{mn}\Gamma^{ab}R_{mn}^{~~ab}\xi_I
= -\frac14R\cdot\xi_I
= 4h\cdot\xi_I~~.
\end{equation}
For round $S^5$ with radius $\ell$, the scalar curvature is
$R=\frac{20}{\ell^2}$ so that $h=-\frac{5}{4\ell^2}$.

The Lagrangian ${\cal L}_\text{SYM}$ for the flat space 
is not invariant under (\ref{susytr}) as it is.
To see this, we take $\delta{\cal L}_\text{SYM}$ and extract the term
containing the auxiliary field $D_{IJ}$,
\begin{equation}
\delta{\cal L}_\text{SYM}\Big|_{{\cal O}(D_{IJ})}
= -2i\text{tr}\big(D^{IJ}\tilde\xi_I\lambda_J\big)~~.
\end{equation}
This can be cancelled by requiring the supersymmetry parameter to
satisfy
\begin{equation}
\tilde\xi_I = t_I^{~J}\xi_J\quad\text{i.e.}\quad
D_m\xi_I= \Gamma_mt_I^{~J}\xi_J~~,
\end{equation}
and by modifying the Lagrangian with the terms
\begin{equation}
{\cal L}'_\text{SYM}= -it^{IJ}\lambda_I\lambda_J+2\sigma t^{IJ}D_{IJ}~~.
\end{equation}
Our convention is $t_I^{~J}\equiv \epsilon^{JK}t_{IK}$. The $SU(2)$
Majorana condition on $\xi_I,\tilde\xi_I$ leads to
\begin{equation}
(t_{IJ})^\ast = \epsilon^{II'}\epsilon^{JJ'}t_{I'J'}~~.
\label{r1}
\end{equation}
Equivalently, $t_I^{~J}$ as a $2\times2$ matrix is a linear sum of
Pauli's matrices with pure imaginary coefficients. One also finds
\begin{equation}
t_I^{~J}t_J^{~K} ~=~ -\frac1{4\ell^2}\delta_I^{~K}~~.
\end{equation}
Thus, we can choose, for example, 
\begin{eqnarray}
t_I^{~J}= \frac{i}{2 l} \sigma_3~~.
\end{eqnarray}
We note that 
$SU(2)$ R-symmetry is broken by nonzero $t_I^{~J}$
to $U(1)$.
We also note that 
the SUSY algebra does not contain dilatation and 
$SU(2)$ R-symmetry except for the unbroken $U(1)$.
This is seen from (\ref{intsym}).

Further computation shows
\begin{equation}
\delta({\cal L}_\text{SYM}+{\cal L}'_\text{SYM}) =
-20it_I^{~J}t_J^{~I}\text{tr}\big(\sigma\epsilon^{IJ}\xi_I\lambda_J\big)
= 10 t^{IJ}t_{IJ}\delta\text{tr}(\sigma^2)~~,
\end{equation}
so that the invariant Lagrangian is 
\begin{eqnarray}
{\cal L}_{S^5} &=&
{\cal L}_\text{SYM}+{\cal L}'_\text{SYM}
-10t^{IJ}t_{IJ}\text{tr}(\sigma^2)
\nonumber \\ &=&
\text{tr}\Big[
\frac12F_{mn}F^{mn}-D_m\sigma D^m\sigma-\frac12D_{IJ}D^{IJ}
+2\sigma t^{IJ}D_{IJ} -10 t^{IJ}t_{IJ}\sigma^2
\nonumber \\ &&\qquad
+i\epsilon^{IJ}\lambda_I\Gamma^mD_m\lambda_J
-\epsilon^{IJ}\lambda_I[\sigma,\lambda_J]
-it^{IJ}\lambda_I\lambda_J
\Big]~~.
\label{lv}
\end{eqnarray}
\vspace{0.2cm}

\paragraph{From Supergravity.}

The rigid SUSY theories on curved space can be obtained from the
corresponding supergravity theory \cite{Festuccia-S}.
The same applies for SUSY gauge theories on $S^5$.
We start from off-shell 5d supergravity coupled to 
Yang-Mills theory \cite{Kugo-Ohashi, Kugo-Ohashi2}.
We give nonzero VEV to the
$SU(2)_R$ triplet auxiliary boson $t_{ij}$ and the metric
in the Weyl multiplet which includes the graviton and gravitino.
This is because the VEV of the scalar will be needed to 
obtain $S^5$ and
the only scalar which appears in the SUSY transformation
of the gravitino is $t_{ij}$.
This also means that the $SU(2)$ R-symmetry should be 
broken to $U(1)$ on $S^5$.
Then, the SUSY transformation of the multiplets vanishes
if 
\begin{eqnarray}
\delta \Psi_{m I} \sim D_m \xi_I +\Gamma_m {t'}_I^{\,J}\xi_J=0~~,
\end{eqnarray}
where $ \Psi_{m I}$ is the gravitino.
Given that the auxiliary field satisfies the reality condition in
(\ref{r1}), this is the Killing spinor subsection if one replaces 
${t'}_I^{\, J} \rightarrow -t_I^{\, J}$.
Accordingly, the SUSY transformation for the vector multiplet is
\begin{eqnarray}
\label{susytr2}
\delta_\xi A'_m  &=& -2 i\epsilon^{IJ}\xi_I\Gamma_m{\lambda'}_J~~, \nonumber \\
\delta_\xi \sigma' &=& 2 i \epsilon^{IJ}\xi_I{\lambda'}_J~~,\nonumber \\
\delta_\xi{\lambda'}_I &=&
\frac14\Gamma^{mn}\xi_I F'_{mn}+\frac{1}{2} \Gamma^m\xi_ID_m \sigma'
-Y_{I}^{\, J} \xi_J~~, \nonumber \\
\delta_\xi Y_{IJ} &=&
i(\xi_I\Gamma^mD_m{\lambda'}_J+\xi_J\Gamma^mD_m{\lambda'}_I)
+  [\sigma',\xi_I{\lambda'}_J+\xi_J{\lambda'}_I]
+i({t'}_I^{\,\,\,\, K} \xi_K {\lambda'}_J+{t'}_J^{\,\,\,\,\, K}
\xi_K{\lambda'}_I)~~, \nonumber \\
\end{eqnarray}
where 
we have used the identity
\begin{eqnarray}
\bar{\xi}^I t^{J}_{\, K} \lambda'^K + 
\bar{\xi}^J t^{I}_{\, K} \lambda'^K 
+2 t^{IJ} \bar{\xi}^K \lambda_K
=\left(  
\bar{\xi}^K t_{K}^{\,\,\,\,\, I} \lambda'^J + 
\bar{\xi}^K t_{K}^{\,\,\,\,\, J} \lambda'^I
\right)~~.
\end{eqnarray}
The action is 
\begin{eqnarray}
{g'}^2 {\cal L}'_{S^5}&=&
\text{tr}\Big[
\frac14F_{mn}F^{mn}-\frac12 D_m\sigma' D^m\sigma'
-Y_{IJ}Y^{IJ}
+4\sigma' {t'}^{IJ}Y_{IJ} -8 {t'}^{IJ} {t'}_{IJ} {\sigma'}^2
\nonumber \\ &&\qquad
+2 i \lambda'_I ( \epsilon^{IJ} \Gamma^mD_m+
{t'}^{IJ} ) \lambda'_J
-2  \epsilon^{IJ} [\lambda'_I ,\lambda'_J] \sigma'
\Big],
\label{lv2}
\end{eqnarray}
where $g'$ is the gauge coupling constant.\footnote{
We renormalized all the fields in order to factor out 
$g'$. Because \cite{Kugo-Ohashi, Kugo-Ohashi2} used the Lorentz signature, 
there might be some ambiguities for the Wick rotation, which
we fix appropriately here.}

One can show from the result above that under the map
\begin{eqnarray}
\sigma &=& - \sigma', \nonumber \\
\lambda_I &=& -2 \lambda'_I, \nonumber \\
D_{IJ} +2 t_{IJ} \sigma&=& 2 Y_{IJ}, \nonumber \\
t_{IJ} &=& -t'_{IJ}, \nonumber \\
{\cal L}_{S^5}  &=& 2 {g'}^2 {\cal L}'_{S^5},
\end{eqnarray}
the SUSY transformation in (\ref{susytr}) and 
the action in (\ref{lv}) are indeed equal to the ones derived from supergravity in (\ref{susytr2}) and (\ref{lv2}).
Using the Chern-Simons term \cite{Kugo-Ohashi},
one can construct the SUSY Chern-Simons term on $S^5$. We have left out the explicit construction in this paper.

For abelian gauge group, FI terms are also SUSY invariant.
On flat $\mathbb R^5$ it is given by
\begin{equation}
{\cal L}_\text{FI}~=~ x^{IJ}D_{IJ},
\end{equation}
where $x^{IJ}$ is an arbitrary $SU(2)_\text{R}$-triplet constant.
On $S^5$, one finds that the FI coupling $x^{IJ}$ has to be
proportional to $t^{IJ}$ and an improvement term must be added.
\begin{equation}
{\cal L}_\text{FI}~=~
t^{IJ}D_{IJ}-6t^{IJ}t_{IJ}\sigma.
\end{equation}
\\

\subsection{Hypermultiplets}\label{sec:Hyper}

In this section, we present the SUSY theories with hypermultiplets. The system of
$r$ hypermultiplets consists of scalars $q^A_I$, fermions $\psi^A$ and
auxiliary scalars $F^A_I$. Here, $I=1,2$ is the $SU(2)$ R-symmetry index
and $A=1,\cdots,2r$. The fields obey the reality conditions
\begin{equation}
(q^A_I)^\ast ~=~ \Omega_{AB}\epsilon^{IJ}q^B_J~~,\quad
(\psi^{A\alpha})^\ast~=~ \Omega_{AB}C_{\alpha\beta}\psi^{B\beta}~~,\quad
(F^A_I)^\ast ~=~ \Omega_{AB}\epsilon^{IJ}F^B_J~~,
\end{equation}
where $\epsilon^{IJ},C_{\alpha\beta}, \Omega_{AB}$ are antisymmetric
invariant tensors of $SU(2)\simeq Sp(1), Spin(5)\simeq Sp(2)$ and
the ``flavor symmetry'' of $r$ free hypermultiplets $Sp(r)$. The
coupling to vector multiplets can be introduced via gauging a subgroup of
$Sp(r)$.
\vspace{0.2cm}

\paragraph{Flat $\mathbb R^5$.}

It is said that one cannot realize off-shell supersymmetry on
hypermultiplets with a finite number of auxiliary fields. Let us review
this by first studying the free theory on $\mathbb R^5$.

It can be easily shown that the Lagrangian
\begin{equation}
{\cal L} ~=~ \epsilon^{IJ}\Omega_{AB}\partial_mq^A_I \partial^mq^B_J
-2i\Omega_{AB}\psi^A\Gamma^m\partial_m\psi^B
\end{equation}
is invariant under the on-shell supersymmetry transformation
\begin{equation}
\delta q^A_I ~=~ -2i\xi_I\psi^A,\quad
\delta \psi^A ~=~ \epsilon^{IJ}\Gamma^m\xi_I\partial_m q^A_J.
\end{equation}
The commutator of two supersymmetries acts on the fields as
\begin{eqnarray}
[\delta_\xi,\delta_\eta]q_I^A &=&
-2i\epsilon^{JK}\xi_J\Gamma^m\eta_K\cdot\partial_mq_I^A~~,\nonumber \\
~[\delta_\xi,\delta_\eta]\psi^A&=&
-2i\epsilon^{IJ}\Gamma^m\eta_I\cdot \xi_J\partial_m\psi^A
-(\xi\leftrightarrow\eta) \nonumber \\
&=&
-i\partial_m\psi^A\cdot\epsilon^{IJ}\xi_I\Gamma^m\eta_J
-i\Gamma^m\partial_m\psi^A\cdot\epsilon^{IJ}\xi_I\eta_J
+i\Gamma^{\ell m}\partial_m\psi^A\cdot\epsilon^{IJ}\xi_I\Gamma_\ell\eta_J
\nonumber \\ &=&
-2i\partial_m\psi^A\cdot\epsilon^{IJ}\xi_I\Gamma^m\eta_J
+\Delta\psi^A~~.
\end{eqnarray}
Here,
\begin{equation}
\Delta\psi^A ~\equiv~
-2i\Gamma^m\partial_m\psi^A\cdot\epsilon^{IJ}\xi_I\eta_J
+2i\eta_I\cdot\epsilon^{IJ}\xi_J\Gamma^m\partial_m\psi^A
-2i\xi_I\cdot\epsilon^{IJ}\eta_J\Gamma^m\partial_m\psi^A,
\end{equation}
and in the last equality we used
\begin{eqnarray}
\lefteqn{
-2i\epsilon^{IJ}\eta_I\cdot \xi_J\Gamma^m\partial_m\psi^A
-(\xi\leftrightarrow\eta)
} \nonumber \\
&=&
-i\partial_m\psi^A\cdot\epsilon^{IJ}\xi_I\Gamma^m\eta_J
-i\Gamma^m\partial_m\psi^A\cdot\epsilon^{IJ}\xi_I\eta_J
-i\Gamma^{\ell m}\partial_m\psi^A\cdot\epsilon^{IJ}\xi_I\Gamma_\ell\eta_J.
\end{eqnarray}
The commutator of two supersymmetries therefore does not precisely close
under a translation by $v^m\equiv 2 \epsilon^{JK}\xi_J\Gamma^m\eta_K$.
The failure terms in $\Delta\psi^A$ are all proportional to the equation
of motion $\Gamma^m\partial_m\psi^A$.

One can try to modify the supersymmetry transformation law by
introducing the auxiliary field $F^A_I$. From dimensional counting
and the symmetries, the only sensible generalization is
\begin{equation}
\delta q^A_I ~\stackrel{?}=~ -2i\xi_I\psi^A,\quad
\delta \psi^A ~\stackrel{?}=~ \epsilon^{IJ}\Gamma^m\xi_I\partial_m q^A_J
+\alpha\epsilon^{IJ}\xi_IF^A_J,
\end{equation}
with an unknown parameter $\alpha$. But whatever the value of $\alpha$ is,
it leads to a failure of the closure of supersymmetry commutators on
$q_I^A$,
\begin{equation}
[\delta_\xi,\delta_\eta]q^A_I ~\stackrel?=~
-2i\epsilon^{JK}\xi_J\Gamma^m\eta_K\cdot\partial_mq_I^A
-2i\alpha\epsilon^{JK}\xi_J\eta_K\cdot F_I^A.
\end{equation}

Thus we will not try to find the transformation law which satisfies
that $[\delta_\xi,\delta_\eta]$ is a translation for any pair $(\xi,\eta)$.
We rather look for the transformation law $\delta$ which satisfies that
$\delta^2$ is a translation for any bosonic $\xi$.\footnote{
Below, we will denote $\delta$ as a fermionic transformation 
generated by a Grassmann-even Killing spinor $\xi$. 
This notation will be used for the localization.}
This property is sufficient for localization.
We propose
\begin{eqnarray}
\delta q^A_I &=& -2i\xi_I\psi^A~~, \nonumber \\
\delta \psi^A &=& \epsilon^{IJ}\Gamma^m\xi_I\partial_m q^A_J
+\epsilon^{I' J'}\check\xi_{I'} F^A_{J'}~~, \nonumber \\
\delta F^A_{I'} &=& 2i\check\xi_{I'}\Gamma^m\partial_m\psi^A~~.
\end{eqnarray}
and an invariant Lagrangian
\begin{equation}
{\cal L} ~=~ \epsilon^{IJ}\Omega_{AB}\partial_mq^A_I \partial^mq^B_J
-2i\Omega_{AB}\psi^A\Gamma^m\partial_m\psi^B
-\epsilon^{I' J'}\Omega_{AB}F^A_{I'} F^B_{J'}~~.
\label{hyperaction}
\end{equation}
Here, $\check\xi_{I'}$ is a constant spinor which satisfies
\begin{equation}
\epsilon^{IJ}\xi_I\xi_J=\epsilon^{I' J'}\check\xi_{I'}\check\xi_{J'}~~,\quad
\xi_I\check\xi_{J'}=0~~,\quad
\epsilon^{IJ}\xi_I\Gamma^m\xi_J+\epsilon^{I' J'}\check\xi_{I'}\Gamma^m\check\xi_{J'}=0~~.
\end{equation}
It looks nontrivial that a spinor $\check\xi_{I'}$ exists for any
choice of $\xi_I$. Therefore, let us prove its existence here.
First, given a pair $(\xi_1,\xi_2)$ of 4-component spinors with the
skew-symmetric inner product
$\xi_1\xi_2\equiv C_{\alpha\beta}\xi_1^\alpha\xi_2^\beta=1$, it is
elementary that one can find two more spinors $\check\xi_1,\check\xi_2$
satisfying
\[
\check\xi_{I'}\xi_J=0~~,\quad
\check\xi_1\check\xi_2=1~~.
\]
Then the tracelessness of $\Gamma^m$ in the basis
$\xi_1,\xi_2,\check\xi_1,\check\xi_2$ gives
\begin{eqnarray}
0~=~\text{Tr}\Gamma^m &=&
\xi_1\Gamma^m\xi_2-\xi_2\Gamma^m\xi_1+\check\xi_1\Gamma^m\check\xi_2
-\check\xi_2\Gamma^m\check\xi_1
\nonumber \\
&=&\epsilon^{IJ}\xi_I\Gamma^m\xi_J+\epsilon^{I' J'}\check\xi_{I'}\Gamma^m\check\xi_{J'}~~.
\end{eqnarray}

We note that 
the action in (\ref{hyperaction}) is invariant under 
the SUSY transformation with
any Killing spinor $\xi_I$ and corresponding $\check \xi_I$.
Thus it gives a 5d $\mathcal{N}=1$ SUSY theory 
with 8 SUSY generators although the commutators between them
include terms other than the (usual) symmetries of the theories.
Furthermore, there is an additional $SU(2)'$ symmetry,
which acts on $I',J'$ indices.


To introduce the coupling to gauge fields and other fields in
the vector multiplet, we need first to introduce the covariant derivative
\begin{equation}
D_m\psi^A ~\equiv~ \partial_m\psi^A-i(A_m)^A_{~B}\psi^B,~\text{etc.}
\end{equation}
Requiring $\Omega_{AB}$ to be gauge-invariant, one finds
$(A_m)_{AB}\equiv\Omega_{AC}(A_m)^C_{~B}$ to be symmetric in the indices $A,B$.
In the following, we introduce the notation $\bar\psi_B\equiv\psi^A\Omega_{AB}$
and suppress the indices $A,B,\cdots$, such that
\begin{eqnarray}
\epsilon^{IJ}\Omega_{AB}D_m q^A_I D^mq^B_J &\equiv&
\epsilon^{IJ}D_m\bar q_I D^m q_J~~,\nonumber \\
\Omega_{AB}\psi^A\Gamma^m(A_m)^B_{~C}\psi^C &\equiv&
\bar\psi\Gamma^mA_m\psi~~,
~~\text{etc.}
\end{eqnarray}
The invariant Lagrangian is
\begin{eqnarray}
{\cal L} &=& \epsilon^{IJ}(D_m\bar q_ID^mq_J -\bar q_I\sigma^2q_J)
-2(i\bar\psi\Gamma^mD_m\psi+\bar\psi\sigma\psi)
\nonumber \\&&
-i\bar q_ID^{IJ}q_J -4\epsilon^{IJ}\bar\psi\lambda_Iq_J
-\epsilon^{I' J'}\bar F_{I'} F_{J'}~~.
\end{eqnarray}
The corresponding SUSY transformation is
\begin{eqnarray}
\delta q_I &=& -2i\xi_I\psi~~, \nonumber \\
\delta\psi &=& \epsilon^{IJ}\Gamma^m\xi_ID_mq_J
+i\epsilon^{IJ}\xi_I\sigma q_J
+\epsilon^{I' J'}\check\xi_{I'} F_{J'}~~, \nonumber \\
\delta F_{I'} &=&
2\check\xi_{I'}(i\Gamma^mD_m\psi+\sigma\psi+\epsilon^{KL}\lambda_Kq_L)~~.
\end{eqnarray}
\vspace{0.2cm}

\paragraph{Five-Sphere.}

Let us first consider the system of free hypermultiplets on $S^5$.
We find that the Lagrangian
\begin{equation}
{\cal L}~=~ \epsilon^{IJ}\Omega_{AB}D_mq^A_I D^mq^B_J
-2i\Omega_{AB}\psi^A\Gamma^mD_m\psi^B
+\frac{15}2\epsilon^{IJ}\Omega_{AB}t^{KL}t_{KL}q^A_Iq^B_J
\end{equation}
is invariant under the on-shell transformation law
\begin{equation}
\delta q^A_I~=~ -2i\xi_I\psi^A~~,\quad
\delta\psi^A~=~ \epsilon^{IJ}\Gamma^m\xi_ID_mq^A_J-3t^{IJ}\xi_Iq_J^A~~.
\end{equation}
Then the unique off-shell extension is given by the Lagrangian
\begin{eqnarray}
{\cal L}&=& \epsilon^{IJ}\Omega_{AB}D_mq^A_I D^mq^B_J
-2i\Omega_{AB}\psi^A\Gamma^mD_m\psi^B
\nonumber \\ &&
+\frac{15}2\epsilon^{IJ}\Omega_{AB}t^{KL}t_{KL}q^A_Iq^B_J
-\epsilon^{I' J'}\Omega_{AB}F^A_{I'} F^B_{J'}~~,
\end{eqnarray}
and the transformation law
\begin{eqnarray}
\delta q^A_I &=& -2i\xi_I\psi^A~~,\nonumber \\
\delta\psi^A &=&
\epsilon^{IJ}\Gamma^m\xi_ID_mq^A_J-3t^{IJ}\xi_Iq_J^A
+\epsilon^{I' J'}\check\xi_{I'}F_{J'}^A~~,\nonumber \\
\delta F^A_{I'} &=&
2i\check\xi_{I'}\Gamma^mD_m\psi^A~~.
\end{eqnarray}
For systems coupled to gauge fields, 
we find that the SUSY invariant Lagrangian is
\begin{eqnarray}\label{hyperla}
{\cal L}_{hyper}&=& \epsilon^{IJ}(D_m\bar q_I D^mq_J-\bar q_I\sigma^2 q_J)
-2(i\psi\Gamma^mD_m\psi+\bar\psi\sigma\psi)
\nonumber \\ &&
-i\bar q_ID^{IJ}q_J-4\epsilon^{IJ}\bar\psi\lambda_Iq_J
+\frac{15}2t^{KL}t_{KL}\epsilon^{IJ}\bar q_Iq_J
-\epsilon^{I' J'}\bar F_{I'}F_{J'}~~,
\end{eqnarray}
with the associated transformation law being
\begin{eqnarray}\label{hypert}
\delta q_I &=& -2i\xi_I\psi,\nonumber \\
\delta\psi &=&
\epsilon^{IJ}\Gamma^m\xi_ID_mq_J
+i\epsilon^{IJ}\xi_I\sigma q_J
-3t^{IJ}\xi_Iq_J
+\epsilon^{I' J'}\check\xi_{I'}F_{J'},\nonumber \\
\delta F_{I'} &=&
2\check\xi_{I'}(i\Gamma^mD_m\psi+\sigma\psi+\epsilon^{KL}\lambda_Kq_L).
\end{eqnarray}

The square of $\delta$ is
\begin{eqnarray}
\delta^2 q_I &=& -i v^m D_mq_I +i \gamma q_I
+R_I^{\,\,\, J} q_J
\nonumber \\
\delta^2\psi &=& -i v^m D_m\psi+i \gamma\psi
+\frac{1}{4} \Theta^{ab} \Gamma^{ab} \psi
\nonumber \\
\delta^2 F_{I'} &=& -i v^m D_m F_{I'}+i \gamma F_{I'}
+{R'}_{I'}^{\,\,\, J'} F_{J'}~~,
\end{eqnarray}
where 
\begin{eqnarray}
v^m &=& \epsilon^{IJ}\xi_I\Gamma^m\xi_J~~, \nonumber \\
\gamma &=& -i\epsilon^{IJ}\xi_I\xi_J\sigma~~, \nonumber \\
R_{IJ} &=& 
3 i (\epsilon^{K L} \xi_K \xi_L) t_{IJ}~~, \nonumber \\
\Theta^{ab} &=&
-2i\epsilon^{IJ} \tilde\xi_I\Gamma^{ab}\xi_J
~~, \nonumber \\
R'_{I' J'}&=& -2i\check\xi_{I'}\Gamma^mD_m\check\xi_{J'}~~.
\end{eqnarray}
Accordingly, $\delta^2$ is a
sum of translation ($v^m$), gauge transformation
($\gamma+i v^m A_m$), 
R-rotation ($R_{IJ}$), Lorentz rotation ($\Theta^{ab}$),
and $SU(2)'$ rotation ($R'_{I' J'}$).
This is consistent with the $\delta^2$ for the vector multiplets.
We can also see that the $R'_{I' J'}$ is indeed in
the $SU(2)'$ from the equation $\epsilon^{I' J'}R'_{I' J'}=0 $
which follows from the definition of $\check{\xi}_{I'}$ 
and the Killing spinor equation.

We can now consider the mass term for the hypermultiplets.
As it is well-known for 4d ${\cal N}=2$ gauge theories,
we can take a decoupling limit of some vector multiplets
to obtain the flavor symmetry and mass terms from
the VEV of the scalar in the vector multiplet.
In our case, we require a constant $m \equiv \langle \sigma \rangle$,
$\langle A_m\rangle=0$, $\langle \lambda \rangle=0$ and 
$\langle D_{IJ} \rangle=-2 t_{IJ} \langle \sigma \rangle$ 
for the unbroken SUSY and the bosonic symmetry.
Accordingly, the mass term is given from $(\ref{hyperla})$ as
\begin{eqnarray}
{\cal L}_{mass}&=& -\epsilon^{IJ}\bar q_I m^2 q_J
-2 \bar\psi m \psi+2 i t^{IJ} \bar q_I m q_J
= \bar q_I \left( -\epsilon^{IJ} m^2+2 i t^{IJ} m \right) q_J 
-2 \bar\psi m \psi. \nonumber \\
\label{hyperm}
\end{eqnarray}
We note that $m$ is an abbreviation for $m_A^{B}$ which should commute
with the remaining gauge symmetry.
We see that the SUSY transformation law in (\ref{hypert}) now depends 
on the mass parameter $m$ even though we are considering 
the off-shell fields
and $\delta^2$ includes
the flavor symmetry generator linear in $m$.
\\

\subsection{Killing Spinors on $S^5$\label{s5killing}}

By now, we have assumed the existence of Killing spinors on $S^5$, $\xi_I$.
In this subsection, we construct them explicitly.
\vspace{0.2cm}

\paragraph{Metric.}

Flat $\mathbb R^5$ and round $S^5$ (with radius $\ell$) have the
metrics
\begin{eqnarray}
ds^2_{\mathbb R^5} &=& \sum_{n=1}^5 dx^ndx^n
~=~ dr^2+ r^2ds^2_{S^4}~~, \nonumber \\
ds^2_{S^5} &=& \ell^2(d\theta^2+\sin^2\theta ds^2_{S^4})~~,
\end{eqnarray}
where $r^2=\sum_{n=1}^5(x^n)^2$.
One can embed a round $S^5$ in flat $\mathbb R^6$, and think of
flat $\mathbb R^5$ which contacts the $S^5$ at its south pole. Then
stereographic projection maps every point on the $S^5$ onto $\mathbb
R^5$ by a line passing through the north pole. It gives the relation
$r = 2\ell\tan\frac\theta2$ and
\begin{equation}
\ell^2d\theta^2 = \frac{dr^2}{(1+\frac{r^2}{4\ell^2})^2}~~,\quad
\ell^2\sin^2\theta = \frac{r^2}{(1+\frac{r^2}{4\ell^2})^2}~~.
\end{equation}
Therefore,
\begin{equation}
ds^2_{S^5}
~=~ \frac{dr^2+r^2ds^2_{S^4}}{(1+\frac{r^2}{4\ell^2})^2}
~=~ \frac{\sum dx_n^2}{(1+\frac{r^2}{4\ell^2})^2}
~=~ \sum_{a=1}^5 e^ae^a~~,
\end{equation}
where $e^a=f\delta^a_n dx^n$ and $f=(1+\frac{r^2}{4\ell^2})^{-1}$.
The spin connection $\omega^{ab}\equiv\sum_c\omega^{ab,c}e^c$ is
determined from the torsion-free condition
\begin{equation}
0 ~=~ de^a+\omega^{ab}e^b ~=~ f^{-2}\partial_nf\delta^n_b\cdot e^be^a
-\omega^{ab,c}e^be^c~~.
\end{equation}
The corresponding solution is
\begin{equation}
\omega^{ab,c}=f^{-2}\partial_nf
(\delta^{ac}\delta^{nb}-\delta^{bc}\delta^{nb})~~.
\end{equation}
\vspace{0.2cm}

\paragraph{Killing spinor equation.}

We first solve the Killing spinor equation without the $SU(2)$ R-index:
\begin{eqnarray}
D_m\Psi =\frac1{2\ell} \Gamma_m \tilde\Psi~~,
\end{eqnarray}
which becomes
\begin{equation}
D_m\Psi ~\equiv~
\partial_m\Psi +\frac12\Gamma^{ab}\delta_m^a\delta^{nb}\partial_n\ln f\Psi
~=~ \frac1{2\ell}f\delta_m^a\Gamma^a\tilde\Psi~~,
\end{equation}
where the coefficient $1/2\ell$ in the right hand side is put for
later convenience, and the Gamma matrices are all coordinate independent.
The above equation can be rewritten as
\begin{equation}
\partial_m(f^{-\frac12}\Psi) ~=~
\frac1{2\ell}f^{\frac12}\delta_m^a\Gamma^a
\big(\tilde\Psi+\ell\delta^n_a\Gamma^a\partial_nf^{-1}\Psi\big)~~.
\end{equation}
The simplest solution is
\begin{equation}
\Psi=f^{\frac12}\Psi_0,\quad
\tilde\Psi = -\frac{\Gamma^ax^a}{2\ell}f^{\frac12}\Psi_0~~.
\end{equation}
One can furthermore find
\begin{equation}
D_m\tilde\Psi ~=~ -\frac1{2\ell}f\delta_m^a\Gamma^a\Psi~~.
\end{equation}

Next, we find the $SU(2)$ Majorana spinor field $\xi_I$ satisfying
\begin{equation}
D_m\xi_I ~=~ t_I^{~J}\Gamma_m\xi_J~~.
\end{equation}
Setting $t_1^{~1}=-t_2^{~2}=\frac i{2\ell},~t_1^{~2}=t_2^{~1}=0$ one
obtains
\begin{eqnarray}
\xi_1 &=& \Big(1+\frac{i\Gamma^ax^a}{2\ell}\Big)f^{\frac12}\Psi_1~~,
\nonumber \\
\xi_2 &=& \Big(1-\frac{i\Gamma^ax^a}{2\ell}\Big)f^{\frac12}\Psi_2~~,
\end{eqnarray}
where $\Psi_1,\Psi_2$ are constant spinors related to each other by
$\Psi_1^\ast=C\Psi_2,~\Psi_2^\ast=-C\Psi_1$.
\vspace{0.2cm}

\paragraph{Bilinears of Killing spinors.}

The scalar bilinear of the Killing spinors takes the value
\begin{equation}
\epsilon^{IJ}\xi_I\xi_J
~=~ 2\xi_1^TC\xi_2
~=~ 2f\Psi_1^T\Big(1+\frac{i(\Gamma^a)^Tx^a}{2\ell}\Big)
C\Big(1-\frac{i\Gamma^ax^a}{2\ell}\Big)\Psi_2
~=~ 2\Psi_2^\dag\Psi_2~~,
\end{equation}
Let us normalize it to unity, $2\Psi_2^\dag\Psi_2=1$.
The vector bilinear takes the form
\begin{eqnarray}
v ~\equiv~
\epsilon^{IJ}\xi_I\Gamma^n\xi_J\frac{\partial}{\partial x^n} &=&
2f\Psi_1^T\Big(1+\frac{i(\Gamma^b)^Tx^b}{2\ell}\Big)
   C\Gamma^n\Big(1-\frac{i\Gamma^cx^c}{2\ell}\Big)\Psi_2
\frac{\partial}{\partial x^n}
\nonumber \\ &=&
2\Psi_2^\dag\Big(1+\frac{i\Gamma^bx^b}{2\ell}\Big)
    \Gamma^a\Big(1-\frac{i\Gamma^cx^c}{2\ell}\Big)\Psi_2
\frac{\partial}{\partial x^a}~~.
\end{eqnarray}
Assuming $\Psi_2$ to be an eigenspinor for $\Gamma^{12}=\Gamma^{34}=i$ and
$\Gamma^5=-1$, the vector bilinear simplifies to the following form
\begin{eqnarray}\label{vecbi}
v &=&
-\frac1\ell\Psi_2^\dag\Gamma^{12}\Psi_2(x^1\partial_2-x^2\partial_1)
-\frac1\ell\Psi_2^\dag\Gamma^{34}\Psi_2(x^3\partial_4-x^4\partial_3)
\nonumber \\ &&
-\left\{
 \Big(1-\frac{x^2}{4\ell^2}\Big)\partial_5
+\frac{x^5x^a}{2\ell^2}\partial_a
\right\}~~.
\end{eqnarray}
We can show, when we regard the $S^5$ as embedded into flat
$\mathbb R^6$, that the above vector bilinear $v$ is a sum of rotations about the 12, 34,
56-planes with an equal angular velocity.
In order to show this, we introduce the Cartesian coordinates $Y_1,\cdots,Y_6$ on
$\mathbb R^6$ to express the round $S^5$ as
\begin{equation}
Y_1^2+\cdots+Y_6^2~=~ \ell^2~~,
\end{equation}
with
\begin{equation}
Y_6= \ell\cos\theta,\quad Y_a= \ell\sin\theta\; \hat r_a~~,
\end{equation}
where $\hat r^a$ is a unit 5-vector. Combining the above with
$x_a=2\ell\tan\frac\theta2\,\hat r_a$, one finds the relation between
the coordinates $(x_1,\cdots,x_5)$ and $(Y_1,\cdots,Y_5)$,
\begin{equation}\label{r6coord}
Y_a=\frac{x_a}{1+\frac{x^2}{4\ell^2}},\quad
\frac\partial{\partial x_a}
=\frac1{1+\frac{x^2}{4\ell^2}}\frac{\partial}{\partial Y_a}
-\frac{x_ax_b}{2\ell^2(1+\frac{x^2}{4\ell^2})^2}
\frac\partial{\partial Y_b}~~.
\end{equation}
Inserting (\ref{r6coord}) into the expression for $v$ in (\ref{vecbi}), one obtains
\begin{equation}
v ~=~ \frac1\ell\Big(
 Y_2\frac\partial{\partial Y_1}
-Y_1\frac\partial{\partial Y_2}
+Y_4\frac\partial{\partial Y_3}
-Y_3\frac\partial{\partial Y_4}
+Y_6\frac\partial{\partial Y_5}
\Big)
\end{equation}
as a vector field on $S^5$ with the coordinate system $(Y_1,\cdots,Y_5)$.
This is a restriction onto the $S^5$ of a vector field $\hat v$ on
$\mathbb R^6$,
\begin{equation}
\hat v ~=~ \frac1\ell\Big(
 Y_2\frac\partial{\partial Y_1}
-Y_1\frac\partial{\partial Y_2}
+Y_4\frac\partial{\partial Y_3}
-Y_3\frac\partial{\partial Y_4}
+Y_6\frac\partial{\partial Y_5}
-Y_5\frac\partial{\partial Y_6}
\Big)~~,
\end{equation}
which is the vector field generating the simultaneous rotations about
the 12, 34, 56-planes by the same angular velocity.
Note that, if we dimensionally reduce the $S^5$ along $v$, we obtain
$\mathbb{C}\mathbb{P}^2$.
\vspace{0.2cm}

\paragraph{Circle fibration over $\mathbb{CP}^2$.}

In order to write the metric on $S^5$ such that the circle fibration structure is
manifest, we introduce the (inhomogeneous) complex coordinates $z^1,z^2$
and an angular coordinate $\vartheta$ to express $Y_1,\cdots,Y_6$ as
\begin{eqnarray}
&&
Y_1+iY_2=Re^{i\vartheta}z^1~~,\quad
Y_3+iY_4=Re^{i\vartheta}z^2~~,\quad
Y_5+iY_6=Re^{i\vartheta}~~,
\nonumber \\
&&
R \equiv \frac\ell{\sqrt{1+|z^1|^2+|z^2|^2}}~~.
\end{eqnarray}
Following the above, the vector field $v$ becomes simply $v=-\frac1\ell\partial_\vartheta$, and
the metric on $S^5$ reads
\begin{eqnarray}
\lefteqn{ ds^2 ~=~ \sum_{i=1}^6 dY_idY_i } \nonumber \\
&=& \ell^2\left[
\left(d\vartheta
    +\frac{i(z^id\bar z^i-\bar z^idz^i)}{2(1+|z^1|^2+|z^2|^2)}\right)^2
+\frac{dz^id\bar z^i}{1+|z^1|^2+|z^2|^2}
-\frac{z^id\bar z^i\cdot dz^j\bar z^j}{(1+|z^1|^2+|z^2|^2)^2}
\right]~~.
\nonumber \\
\end{eqnarray}
With the notation
\begin{eqnarray}
&&
ds^2 ~=~ \ell^2\left[
(d\vartheta+V)^2+2g_{i\bj}dz^i d\bar z^j
\right],\quad
V~=~ V_idz^i+V_\bi d\bar z^i,
\nonumber \\ &&
g_{i\bj} ~=~
\frac12\partial_i\bar\partial_{\bj}\ln(1+|z^1|^2+|z^2|^2)~~,
\end{eqnarray}
we have
\begin{equation}
dV ~=~ 2ig_{i\bj}dz^i\wedge d\bar z^j~~.
\end{equation}

If we use the above metric on $S^5$, a contravariant vector $X$ has
components $X^1,X^2,X^{\bar 1},X^{\bar 2}$ and $X^\vartheta$. The inner
product of contravariant vectors are
\begin{equation}
G_{mn}X^mY^n~=~
\ell^2\left[
 (X^\vartheta+V_iX^i+V_\bi X^\bi)
 (Y^\vartheta+V_iY^i+V_\bi Y^\bi)
+g_{i\bj}(X^iY^\bj+X^\bj Y^i)\right]~~,
\end{equation}
with the component of the metric $G_{mn}$ being
\begin{eqnarray}
&&
G_{\vartheta\vartheta}=\ell^2~~,\quad
G_{\vartheta i}=\ell^2V_i~~,\quad
G_{\vartheta\bi}=\ell^2V_\bi~~,
\nonumber \\ &&
G_{i\bj}=\ell^2(g_{i\bj}+V_iV_\bj)~~,\quad
G_{ij}=\ell^2V_iV_j~~,\quad
G_{\bi\bj}=\ell^2V_\bi V_\bj~~.
\end{eqnarray}
The inverse metric has components
\begin{eqnarray}
&&
G^{\vartheta\vartheta}=\ell^{-2}(1+2g^{i\bj}V_iV_\bj)~~,\quad
G^{\vartheta i}=-\ell^{-2}g^{i\bj}V_\bj~~,\quad
G^{\vartheta\bi}=-\ell^{-2}g^{j\bi}V_j~~,
\nonumber \\ &&
G^{i\bj}=\ell^{-2}g^{i\bj}~~,\quad
G^{ij}= G^{\bi\bj}=0~~,
\end{eqnarray}
where $g^{i\bj}$ is the inverse metric on the base $\mathbb{CP}^2$, namely
$g_{i\bj}g^{k\bj}=\delta_i^k$.
\\

\vspace{1cm}
\section{Localization}

In this section, we apply localization
to the 5d gauge theories.
\vspace{0.2cm}

\paragraph{Vector Multiplets.}

Let us first concentrate on the vector multiplets
and choose a Killing spinor $\xi_I$.
Denoting the corresponding SUSY transformation as $\delta$, one notes that
$\delta^2$ is a combination of the transformation generated by $v^m$
and an $U(1)$ R- and Lorentz transformation.
Assuming that the transformation $\delta$ is the quantum mechanical symmetry,
we obtain
\begin{eqnarray}
\frac{d}{dt} \langle {\cal O}_1 {\cal O}_2 \cdots {\cal O}_n
e^{-t \delta I} \rangle =0~~,
\end{eqnarray}
where ${\cal O}_i$ and $I=\int_{S^5} V$ are assumed to satisfy $\delta {\cal O}_i=0$
and 
\begin{eqnarray}
\delta^2 I=0~~. 
\end{eqnarray}
Here, we also assume that $\delta I$ is (real) positive definite
in the path-integral.
Accordingly, by taking $ t\rightarrow \infty$,
the path-integral is localized on the constraint $\delta I=0$
with the one loop determinant for the regulator action being $-\delta I$.

To explain our choice of the regulator Lagrangian, we recall
\begin{equation}
\delta\lambda_I ~=~
-\frac12\Gamma^{mn}\xi_IF_{mn}+\Gamma^m\xi_ID_m\sigma
+\xi_J(D_{KI}+\sigma t_{KI})\epsilon^{JK}~~.
\end{equation}
In this section, we take $\xi_I$ as Grassmann-even such that $\delta$ is the fermionic transformation.
Of course, this is the symmetry of the action because 
it is linear in $\xi_I$.
Note that the commutator $[\delta_\xi, \delta_\xi ]$ 
becomes $2 \delta^2$.
For $ \delta^2$,
the right-hand side of the commutators in $(\ref{com})$ are unchanged,
but the parameters become 
\begin{eqnarray}
v^m &=& \epsilon^{IJ}\xi_I\Gamma^m\xi_J~~, \nonumber \\
\gamma &=& -i\epsilon^{IJ}\xi_I\xi_J\sigma~~, \nonumber \\
\rho &=& 0~~,  \nonumber \\
R_{IJ} &=& -3i(\xi_I\tilde\xi_J+\xi_J\tilde\xi_I
             )=3 i (\epsilon^{K L} \xi_K \xi_L) t_{IJ}~~, \nonumber \\
\Theta^{ab} &=&
-2i\epsilon^{IJ} \tilde\xi_I\Gamma^{ab}\xi_J
~~.
\label{intsym2}
\end{eqnarray}

To make the SYM Lagrangian positive definite, we notice that the path
integration contours for $\sigma$ and $D$ have to be rotated by 90
degrees, which implies that they are regarded as purely imaginary.\footnote{
Here, the SUSY action and the SUSY transformation
are written in terms of $\lambda_I, \sigma, D_{IJ}$, which are holomorphic.
Accordingly, the action is SUSY invariant for any choice of the contour.
This is clear because we have not used 
$\lambda_I^\dagger, \sigma^\dagger, D^\dagger_{IJ}$, 
in the Lagrangian and in the SUSY transformation.
The choice here corresponds to, for example,
$\sigma^\dagger=-\sigma$, which is not the relation
$\sigma^\dagger=\sigma$
originally assumed.}
Accordingly, the complex conjugate of the above formula is
\begin{equation}
(\delta\lambda_I)^\ast ~=~
-\frac12\epsilon^{II'}C\Gamma^{mn}\xi_{I'}F_{mn}
-\epsilon^{II'}C\Gamma^m\xi_{I'}D_m\sigma
-\epsilon^{II'}C\xi_J(D_{KI'}+\sigma t_{KI'})\epsilon^{JK}~~.
\end{equation}
Its transpose is
\begin{equation}
(\delta\lambda_I)^\dagger ~=~
+\frac12\epsilon^{I'I}F_{mn}\xi_{I'}C\Gamma^{mn}
-\epsilon^{I'I}D_m\sigma\xi_{I'}C\Gamma^m
-\epsilon^{I'I}(D_{KI'}+\sigma t_{KI'})\xi_JC\epsilon^{JK}~~.
\end{equation}
We then take the regulator Lagrangian of the form $\delta V$, with
\begin{eqnarray}
V &=& \text{tr}\big[(\delta\lambda)^\dagger \lambda\big]
\nonumber \\
 &=& \text{tr}\left[
\frac12\epsilon^{IJ}\xi_I\Gamma^{mn}\lambda_J F_{mn}
-\epsilon^{IJ}\xi_I\Gamma^m\lambda_JD_m\sigma
-\epsilon^{IJ}\xi_K\lambda_J(D_{LI}+2\sigma t_{LI})\epsilon^{KL}
\right]
\end{eqnarray}
and $\xi_I$ being Grassmann-even.\footnote{
We can think of the right-hand side as the definition of $I$
and forget about the definition of $\delta (\lambda_I)^\dagger$.}
One should note that
\begin{eqnarray}
\delta^2 \int_{S^5} V=0~~,
\label{e1}
\end{eqnarray}
which can be shown as follows.
The $\delta^2$ is the bosonic symmetry transformation
of (\ref{intsym2}), however, $\xi_I$, whcih is not a field,  
does not transform under the transformation.
Since all indices are properly contracted in $V$,
(\ref{e1}) is correct if $\xi_I$ would transform under the symmetry
as its indices indicate.
This is possible if $\xi_I$ is invariant under this.
Indeed, the Jacobi identity $[\delta^2,\delta]=0$
implies that $\xi_I$ is invariant. This is
because by decomposing $\delta$ to a spinor and to $SU(2)$ components as
$\delta=\xi_{\alpha I} \delta^{\alpha I}$, we see that 
$[\delta^2,\delta]=\delta_{\xi'}$ where $\xi'$ is the transformation 
of $\xi$ by the bosonic symmetry.
We can also show the invariance explicitly by using the identity
$t^{KJ} w_{KJ}^{ab} \Gamma_{ab} \xi_I=-4 t_I^{\,\,\,\, J} \xi_J$ 
followed from the Fierz identity.

$\delta V$ consists of a collection of purely bosonic terms and
terms bilinear in the fermion. The purely bosonic terms read
\begin{eqnarray}
\delta V|_\text{bos} &=&
\text{tr}\bigg[\frac12F^{mn}F_{mn}-D_m\sigma D^m\sigma
-\frac12(D_{IJ}+2\sigma t_{IJ})(D^{IJ}+2\sigma t^{IJ})
\nonumber \\ && ~~~
-\frac14v_p\varepsilon^{klmnp}F_{kl}F_{mn}\bigg]~~.
\end{eqnarray}
Using $v_mv^m=1$ (which is derived below), one can complete the square such that
\begin{eqnarray}
\delta V|_\text{bos} &=&
\text{tr}\bigg[
\frac14(F_{mn}-\frac12\epsilon_{mnpqr}v^pF^{qr})
(F^{mn}-\frac12\epsilon^{mnstu}v_sF_{tu})
+\frac12 (v^pF_{pm})(v_qF^{qm})
\nonumber \\ && ~~~
-D_m\sigma D^m\sigma -\frac12(D_{IJ}+2\sigma t_{IJ})(D^{IJ}+2\sigma t^{IJ})
\bigg]~~.
\end{eqnarray}
This is indeed positive definite 
(by definition) for our choice of the contour.
The saddle point condition $\delta V|_{bos}=0$ is therefore
\begin{equation}
F_{mn}=\frac12\epsilon_{mnpqr}v^pF^{qr}~~,\quad
v^mF_{mn}=0~~,\quad
D_m\sigma=0~~,\quad
D_{IJ}+2\sigma t_{IJ}=0~~,
\label{saddlept}
\end{equation}
where the first equation implies the second equation.
This kind of instanton equations was studied in \cite{Harland}.

Below, we derive the equality $v_mv^m=1$. This follows from a stronger
equality
\begin{equation}
\Gamma_m\xi_I\cdot v^m ~=~ \xi_I~~.
\label{key-eq}
\end{equation}
To show this, we look into the following Fierz identities\footnote{Note the sign
difference from the previous formula due to the fact that we are here dealing with
Grassmann even spinors.}
\begin{eqnarray}
\lefteqn{
\xi_I ~=~ \xi_I\epsilon^{JK}(\xi_J\xi_K)
} \nonumber \\
&=&
\frac14\xi_K\epsilon^{JK}(\xi_J\xi_I)
+\frac14\Gamma_\ell\xi_K\epsilon^{JK}(\xi_J\Gamma^\ell\xi_I)
-\frac18\Gamma_{\ell m}\xi_K\epsilon^{JK}(\xi_J\Gamma^{\ell m}\xi_I)~~,
\nonumber \\
\lefteqn{
\Gamma_n\xi_I\cdot v^n ~=~ \Gamma_n\xi_I\epsilon^{JK}(\xi_J\Gamma^n\xi_K)
}
\nonumber \\
&=&
\frac14\xi_K\epsilon^{JK}(\xi_J\Gamma^n\Gamma_n\xi_I)
+\frac14\Gamma_\ell\xi_K\epsilon^{JK}(\xi_J\Gamma^n\Gamma^\ell\Gamma_n\xi_I)
-\frac18\Gamma_{\ell m}\xi_K\epsilon^{JK}(\xi_J\Gamma^n\Gamma^{\ell m}\Gamma_n\xi_I)
\nonumber \\
&=&
\frac54\xi_K\epsilon^{JK}(\xi_J\xi_I)
-\frac34\Gamma_\ell\xi_K\epsilon^{JK}(\xi_J\Gamma^\ell\xi_I)
-\frac18\Gamma_{\ell m}\xi_K\epsilon^{JK}(\xi_J\Gamma^{\ell m}\xi_I)~~.
\end{eqnarray}
Using
\begin{equation}
\xi_J\xi_I=-\frac12\epsilon_{JI}~~,\quad
\xi_J\Gamma^n\xi_I=-\frac12\epsilon_{JI}v^n~~,
\end{equation}
one finds
\begin{eqnarray}
\xi_I
&=&
\frac18\xi_I
+\frac18\Gamma_\ell\xi_I\cdot v^\ell
-\frac18\Gamma_{\ell m}\xi_K\epsilon^{JK}(\xi_J\Gamma^{\ell m}\xi_I)~~,
\nonumber \\
\Gamma_n\xi_I\cdot v^n
&=&
\frac58\xi_I
-\frac38\Gamma_\ell\xi_I\cdot v^\ell
-\frac18\Gamma_{\ell m}\xi_K\epsilon^{JK}(\xi_J\Gamma^{\ell m}\xi_I)~~.
\end{eqnarray}
By taking the difference between the above two equations, one finds the desired
equality in (\ref{key-eq}).

Next, we show that $\delta_\xi\lambda_I=0$ follows from the saddle point condition in
(\ref{saddlept}). Recall
\begin{equation}
\delta_\xi\lambda_I ~=~
-\frac12\Gamma^{mn}\xi_I F_{mn}+\Gamma^m\xi_I D_m\sigma
+\xi_J(D_{KI}+2\sigma t_{KI})\epsilon^{JK}~~.
\end{equation}
Assuming (\ref{saddlept}), all the terms on the right hand side except
for the first one vanish. To show that the first term also
vanishes, we notice
\begin{eqnarray}
\Gamma^{mn}\xi_IF_{mn}
&=& \frac12\Gamma^{mn}\xi_I\epsilon_{mnpqr}F^{pq}v^r
~=~ -\Gamma^{pqr}\xi_I F_{pq}v_r
\nonumber \\
&=& -(\Gamma^{pq}\Gamma^r-\Gamma^pg^{qr}+\Gamma^qg^{pr})\xi_IF_{pq}v_r
~=~ -\Gamma^{pq}\xi_IF_{pq}~~.
\end{eqnarray}
Therefore, $\Gamma^{mn}\xi_IF_{mn}$ vanishes. Note that,
in the second equality, we use $\Gamma^{12345}=1$ leading to
$\Gamma^{mn}\epsilon_{mnpqr}=-2\Gamma_{pqr}$. We also use
(\ref{key-eq}) in the fourth equality. Since $\delta_\xi I|_\text{bos}$
vanishes by construction if $\delta_\xi\lambda_I=0$, it follows that
(\ref{saddlept}) and $\delta_\xi\lambda_I=0$ are completely equivalent.

Let us now consider the saddle point equation in (\ref{saddlept}).
We recall that  $v^m F_{mn}$ means a translation (Lie derivative) 
with $v^m$ and a gauge transformation with $v^m A_m$ of $A_n$.
Thus, if we can take $v^m A_m=0$ gauge,
the condition $v^m F_{mn}=0$ means
$A_n$ is constant in the $v^m$ direction.
Accordingly, we can think of the gauge field as being only on $\mathbb{CP}^2$. It should be an instanton solution which follows from 
the condition $ F_{mn}=\frac12\epsilon_{mnpqr}v^pF^{qr}$.
If, for example, a Wilson line for $v^m A_m0$ does not vanish,
we can not take the gauge. In this case, the saddle points are 
a combination of the Wilson line and the instantons. 
Therefore, we conclude that 
the path-integral is reduced to an
integration over a generalization
of instantons on $\mathbb{CP}^2$
and the covariantly constant $\sigma$ on it.
Needless to say, it is important to carry out explicitly
this integral with the one loop determinant factor 
and saddle point action. We hope to return to this problem 
in the near future.
\vspace{0.2cm}

\paragraph{Hypermultiplets.}

Finally, we consider the localization of the hypermultiplets.
If
\begin{eqnarray}
\delta\psi &=&
\epsilon^{IJ}\Gamma^m\xi_ID_mq_J
+i\epsilon^{IJ}\xi_I (\sigma+m) q_J
-3t^{IJ}\xi_Iq_J
+\epsilon^{I' J'}\check\xi_{I'}F_{J'}~~,
\end{eqnarray}
then its complex conjugate should be (before rotating the integration
contours for some variables),
\begin{eqnarray}
(\delta\psi)^\ast &=&
\Omega C\Big(
\epsilon^{IJ}\Gamma^m\xi_ID_mq_J
+i\epsilon^{IJ}\xi_I (\sigma+m) q_J
-3t^{IJ}\xi_Iq_J
+\epsilon^{I' J'}\check\xi_{I'}F_{J'}\Big)~~.
\end{eqnarray}
For positivity of the action of the hypermultiplets
with the mass term, 
we have assumed that
$F$ is ``pure imaginary'', $q$ is ``real'' and
the complex conjugate of $m$ is the same as the one for $\sigma$.
With the rotation of the contours for $\sigma, D_{IJ}, F_{J'}$
(and $m$) taken into account, this is modified to
\begin{eqnarray}
(\delta\psi)^\ast &=&
\Omega C\Big(
\epsilon^{IJ}\Gamma^m\xi_ID_mq_J
-i\epsilon^{IJ}\xi_I (\sigma+m) q_J
-3t^{IJ}\xi_Iq_J
-\epsilon^{I' J'}\check\xi_{I'}F_{J'}\Big)~~.
\end{eqnarray}
By taking its transpose one finds
\begin{eqnarray}
(\delta\psi)^\dagger &=&
\epsilon^{IJ}\xi_IC\Gamma^mD_mq_J\Omega
+i\epsilon^{IJ}\xi_IC q_J\Omega(\sigma+m)
-3t^{IJ}\xi_ICq_J\Omega
-\epsilon^{I' J'}\check\xi_{I'}CF_{J'}\Omega~~. \nonumber\\
\end{eqnarray}
The regulator Lagrangian for the localization will be  $\delta V_{hyper}$
where
\begin{eqnarray}
V_{hyper} &=&  (\delta\psi)^\dagger \psi~~.
\end{eqnarray}
Then, the bosonic part of 
the regulator Lagrangian is 
$\delta V_{hyper}|_{bos}= (\delta\psi)^\dagger \delta \psi$
which becomes
\begin{eqnarray}
\delta V_{hyper} &=&
\frac12 \epsilon^{IJ} 
D_m \bar{q}_I D_m q_J
+3 v^m t^{IJ} \bar{q}_I D_m q_J
+\frac{9}{4} t^{IJ} t_{IJ} \epsilon^{KL} \bar{q}_K q_L
\nonumber \\
&& 
+w^{mn}_{\,\,\, IJ} D_m \bar q^I D_n q^J
-\frac12\epsilon^{IJ}\bar q_I(\sigma+m)^2q_J
-\frac12 \epsilon^{I' J'} \bar{F}_{I'}F_{J'}~~.
\end{eqnarray}
Above, we have defined
\begin{eqnarray}
w^{mn}_{\,\,\, IJ} \equiv \xi_{I} \Gamma^{mn} \xi_J~~,
\end{eqnarray}
which satisfies $w^{mn}_{\,\,\, IJ}=w^{mn}_{\,\,\, JI}
=-w^{nm}_{\,\,\, IJ}$.
Using the identities, we show from
the Fierz identities that
\begin{eqnarray}
\delta V_{hyper} &=&
\frac12 
\epsilon^{IJ} 
\left( 
v^m D_m \bar{q}_I
-3  t_I^{\,\,\, K} \bar{q}_K
\right)
\left( v^n D_n q_J
-3  t_J^{\,\,\, L} q_L 
\right)
\nonumber \\
&& 
+\frac{1}{8} 
\epsilon^{KL} 
\left( 
D^p \bar{q}_K-v^p (v^q D_q \bar{q}_K) 
+2 w^{pm \,\, I}_{\,\,\,\, K} D_m \bar{q}_I
\right)
\left( 
D_p q_L-v_p (v^r D_r q_L) 
+2 w_{pnL}^{\,\,\,\,\,\,\,\,\,\,\, J} D^n q_J
\right)
\nonumber \\
&& 
-\frac12\epsilon^{IJ}\bar q_I(\sigma+m)^2q_J
-\frac12 \epsilon^{I' J'} \bar{F}_{I'}F_{J'}~~,
\label{hsp}
\end{eqnarray}
where
each term is positive definite for our choice of the contour.
Therefore, the conditions for the saddle points are
\begin{eqnarray}
&& C_K \equiv v^m D_m q_K
-3  t_K^{\,\,\, L} q_L =0~~,  \nonumber \\
&& {C'}_{p K} \equiv  D_p q_K-v_p (v^r D_r q_K) 
+2 w_{pnK}^{\,\,\,\,\,\,\,\,\,\,\, J} D^n q_J =0~~,
\nonumber \\
&& (\sigma+m) q_J=0, \,\,\,\,\,\,\,\,\, 
F_{J'} =0~~.
\end{eqnarray}
Furthermore, in the Coulomb branch, where $ (\sigma+m)$
does not have zero eigenvalue, the conditions are trivial;
$q_J=0$ and $F_{J'} =0$.
On the other hand, 
a Higgs or mixed branch would exist
if $ (\sigma+m)$ has zero eigenvalues.\footnote{
However, it is possible that there are no solutions 
of $C_K={C'}_{pK}=0$ on $S^5$.}

Let us derive some identities for 
$w^{mn}_{\,\,\, IJ}$.
Multiplying $\xi_L \Gamma^{p_1 p_2 \cdots}$ with the Fierz identity,
we have the following:
\begin{eqnarray}
\label{i1}
0 &=& 
-w_{mnIJ} \, w^{mn}_{\,\,\,\,\,\, KL}
+\epsilon_{IJ} \epsilon_{KL}+2 \epsilon_{IL} \epsilon_{JK}~~, \\
\label{i2}
0 &=& 2 \left(
\epsilon_{IJ} \epsilon_{KL}+2 \epsilon_{IL} \epsilon_{JK}
\right) v^p
+2 v_m \left(
\epsilon_{IJ} w^{pm}_{\,\,\,\,\,\,\, KL} - \epsilon_{KL}
w^{pm}_{\,\,\,\,\,\,\, IJ} 
\right) 
\nonumber \\ && 
+ \epsilon^{pmnqr} w_{mn IJ} w_{qrKL}~~, \\
\label{i3}
0 &=& 8\epsilon_{JK} w^{pq}_{\,\,\,\,\,\,\, LI}
-2\epsilon_{JI} w^{pq}_{\,\,\,\,\,\,\, LK}
-2\epsilon_{LK} w^{pq}_{\,\,\,\,\,\,\, JI}
\nonumber \\ && 
+\epsilon_{JI} v_m \epsilon^{pqmrs} w_{rs LK}
+\epsilon_{LK} v_m \epsilon^{pqmrs} w_{rs JI}
\nonumber \\ && 
-4\left(
w^{qn}_{\,\,\,\,\,\,\, JI} w^{p}_{\,\,\,\, nLK}
-w^{pn}_{\,\,\,\,\,\,\, JI} w^{q}_{\,\,\,\, nLK}
\right)~~.
\end{eqnarray}
Then, by applying $\epsilon^{IJ}$ to the identities 
(\ref{i2}) and (\ref{i3}), we obtain
\begin{eqnarray}
\label{i4}
&& 0=v_m w^{pm}_{\,\,\,\,\,\,\, KL}~~, \\
\label{i5}
&& 0=2 w^{pq}_{\,\,\,\,\,\,\, IJ}
+ v_m \epsilon^{pqrsm}
w_{rs IJ}~~.
\end{eqnarray}
In addition, by applying $\epsilon^{LI}$ to the identities 
(\ref{i1}), (\ref{i2}) and (\ref{i3}), we obtain
\begin{eqnarray}
\label{i6}
&& 0= w^{mn}_{\,\,\,\,\,\,\, KI} w_{mn J}^{\,\,\,\,\,\,\,\,\,\,\,\, I}
+3 \epsilon_{JK}~~, \\
\label{i7}
&& 0=6 \epsilon_{JK} v^p
-\epsilon^{pmnqr} w_{mn IJ} w_{qr K}^{\,\,\,\,\,\,\,\,\,\,\,\, I}~~, \\
\label{i8}
&& 0= -2 w^{qp}_{\,\,\,\,\,\,\, JK}
+ \epsilon^{IL} \left(
w^{qn}_{\,\,\,\,\,\,\, IJ} w^{p}_{\,\,\,\, nLK}
+w^{qn}_{\,\,\,\,\,\,\, IK} w^{p}_{\,\,\,\, nLJ}
\right)~~.
\end{eqnarray}
By taking the square of 
the first term and the second term of (\ref{i5}), 
we have the following identity
\begin{equation}
\epsilon^{JL}
(w^{pm}_{~~~~IJ}w^{q}_{~mKL}+w^{qm}_{~~~~IJ}w^{p}_{~mKL})
= \frac32\epsilon_{IK}(g^{pq}-v^pv^q)\,~~.
\end{equation}
With this and (\ref{i8}) we find
\begin{eqnarray}
\label{i9}
&& 
\epsilon^{KL} 
w^{pm}_{\,\,\,\,\,\,\, IK} w^{q}_{\,\,\,\, mJL}
=w^{pq}_{\,\,\,\,\,\,\, IJ}
+\frac{3}{4} \left( g^{pq} - v^p v^q\right) \epsilon_{IJ}~~.
\end{eqnarray}
Using the identities we derived, 
we can show the form of the bosonic part in (\ref{hsp}) 
where the positive definiteness is manifest.

We can also show that $\delta \psi=0$ is indeed satisfied
by the saddle point condition.
It is easy to see that 
$\xi_K \delta \psi=0$ and 
$\xi_K \Gamma^p \delta \psi=0$ on the saddle points.
We also find the identity:
$\xi_K \Gamma^{pq} \delta \psi
+v^q (\xi_K \Gamma^p \delta \psi)- v^p (\xi_K \Gamma^q \delta \psi)
-2 (\xi_I \delta \psi) w^{pq}_{KJ} \epsilon^{IJ}  =0$.
Therefore, we have indeed shown that 
$\delta \psi=0$ is indeed satisfied on the saddle points.
\\

\section{Conclusion}

In this paper, we have constructed 5d SUSY gauge
theories on the five-sphere with vector and hypermultiplets.
We have shown that with the localization terms
the path-integral can be restricted to 
an integration over a generalization of instantons 
on $\mathbb{CP}^2$ and the covariantly constant Coulomb moduli.
It is interesting that
instantons in 4d appear in 5d SUSY gauge theories on $S^5$.
If we regard the 5d theory as a compactification
of a 6d conformal field theory,
the 4d instanton in the former 
can be interpreted as a Kaluza-Klein particle of the latter.
Then, the choice of the Killing spinor, which determines $v^m$, 
corresponds to a choice of a Wick rotated time direction.

It is of great interest to study further the appearance of instantons on $\mathbb{C}\mathbb{P}^2$ in the context of 5d SUSY theories on $S^5$. Evaluating the integration over the moduli space of instantons and hence finding the localized path-integral are key steps to take in future work. In addition, of great interest is the study of SUSY gauge theories on deformed $S^5$ in line with the work in \cite{Hama:2011ea} for the case of the squashed three-sphere.

\section*{Acknowledgements}

K. H. would like to thank the Perimeter Institute
for hospitality. R.-K. S. would like to thank the Yukawa Institute at Kyoto University for hospitality, and the Global COE program for support.
S. T. would like to thank K. Sakai and M. Taki for helpful discussions.
The work of S. T. is partly supported by the Japan Ministry of Education,
Culture, Sports, Science and Technology (MEXT), and by the Grant-in-Aid
for the Global COE program ``The Next Generation of Physics, Spun from
Universality and Emergence'' from the MEXT.

\vspace{1cm}

\noindent
{\bf Note added}: 

As this article neared completion,
we became aware of the preprint \cite{Kallen-Z}
in which topological SUSY 
gauge theories on 5d manifolds with circle fibration
structure, including spheres, are constructed
following the 3d case \cite{Kallen}.
Our 5d SUSY gauge theories on $S^5$ 
with the terms needed for localization would
coincide with their topological one if 
we ignore the original action. This would be 
equivalent to taking the (formal) strong coupling limit.

\end{document}